\documentclass[aps,twocolumn,groupedaddress, showpacs,prb]{revtex4}
\usepackage{graphicx}%

\begin{document}
\bibliographystyle{apsrev}


\title{Zero-loss/deflection map analysis}


\author{P. Fraundorf}
\email[]{pfraundorf@umsl.edu}
\author{K. Pisane}
\author{E. Mandell}
\author{R. Collins}
\affiliation{Physics \& Astronomy/Center for NanoScience, U. Missouri-StL (63121)}


\date{\today}

\begin{abstract}

Experimental plots of the fraction of detected electrons removed from the zero-loss peak, versus the fraction of incident electrons scattered outside of the objective aperture, can serve as a robust fingerprint of object-contrast in an energy filtered transmission electron microscope (EFTEM).  Examples of this, along with the first in a series of models for interpreting the resulting patterns, were presented at the August 2010 meeting of the Microscope Society of America meeting in Portland, Oregon, and published in {\em Microscopy and MicroAnalysis} {\bf 16}, Supplement 2, pages 1534-1535 by Cambridge University Press.

\end{abstract}
\maketitle

\tableofcontents

\section{Introduction}
\label{sec:Intro}



Outside of microscopy the relationship between energy lost, and momentum transferred, is a subject that has found considerable application in the past 2000 years. Only recently has it become possible for electron microscopes to routinely deliver quantitative information on energy lost as well as momentum transfer. 

One simple and robust way to put this type of information to use is to group image-pixels according to the fraction of detected electrons removed from the zero-loss peak\cite{egerton96} versus the fraction of incident electrons scattered outside the objective aperture. These plots show new promise as a robust tool for quantitative analysis of intensities in transmission electron microscopes able to form energy-filtered images. 

\section{Zero-loss versus deflected intensity}
\label{sec:fractionPlots}

\begin{figure}
\includegraphics[scale=0.75]{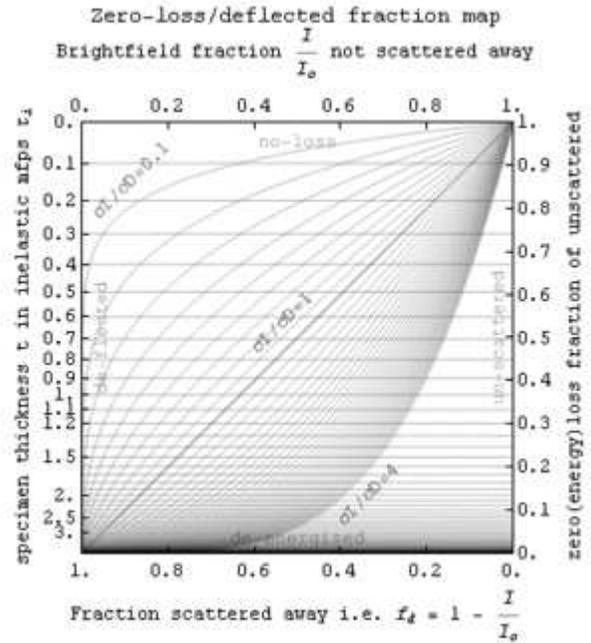}%
\caption{Fraction of detected electrons with zero-loss, versus the fraction of incident electrons undeflected.}
\label{fig1}
\end{figure}

Figure \ref{fig1} shows a map on which anyone can plot experimental observations of specimen brightness I/Io in a brightfield transmitted electron image (x-axis), versus the fraction of detected electrons (y-axis) that did not lose appreciable energy (e.g. more than a few ppm). 
Lateral coherence effects aside, both of these are simple to understand experimental quantities between 0 and 1 for a given scattering experiment. Holes in the specimen by definition plot at the (1,1) point in the upper right corner, while regions too thick to permit electrons through plot at (0,0) in the lower left. An electron energy filter or spectrometer is needed to characterize the y-position of a specimen region on the plot, but not to obtain data on the x-position. 

\section{Application examples}
\label{sec:examples}

\begin{figure}
\includegraphics[scale=0.75]{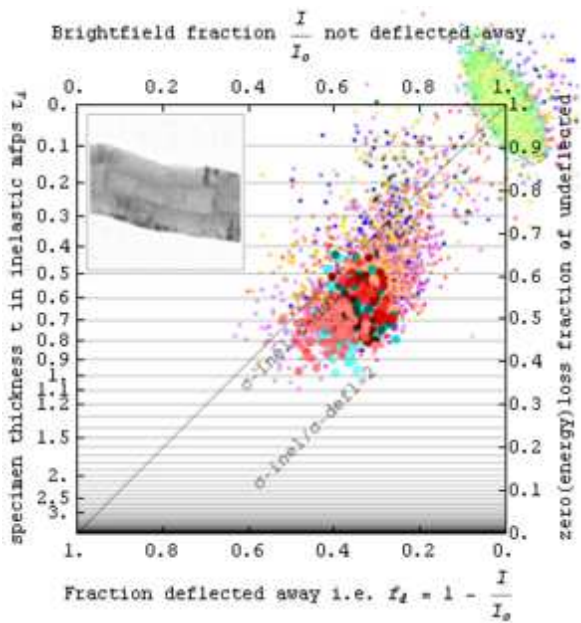}%
\caption{Bamboo C-nanorod: Small dots cover the whole image; large dots cyan-red run left-right through the center.}
\label{fig2}
\end{figure}

\subsection{multiwalled tube}
\label{sec:nanotube}

Figure \ref{fig2} shows a zero-loss/deflection map for pixels from a brightfield/zero-loss image pair taken of a bamboo-type multiwalled nanotube. This specimen allows one to examine cross-sections in specimens of known thickness and (002) orientation with respect to the electron beam. 

\subsection{presolar graphene}
\label{sec:presolar}

Figure 3 shows the inelastic cross-section (and perhaps density as well) of unlayered-graphene cores\cite{Mandell2007} to be about 80\% that of the graphite rim in microtomed presolar core-rim graphite onions\cite{Fraundorf2002} from the Murchison meteorite\cite{Bernatowicz1996}.  Zero-loss/deflection maps of multiple onions support this conclusion. 

\begin{figure*}
\includegraphics[scale=0.75]{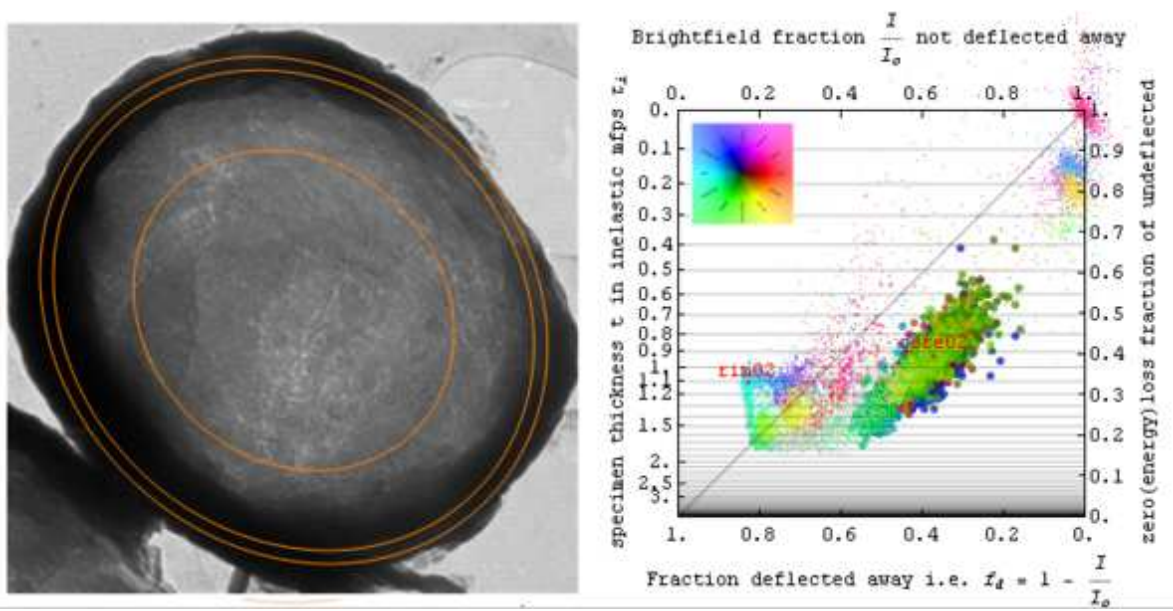}%
\caption{Left panel - brightfield image of presolar slice "Vivian"; Right panel - core region dots are large, rim region dots are medium, transition region and support film dots are small; Except for the gray transition region dots, dot color denotes location in the image using the color template in the upper left.}
\label{fig3}
\end{figure*}



\begin{acknowledgments}
Thanks to Howard Berg of Danforth Plant Sciences Center for use of their Zeiss Leo 912AB
EFTEM, and for specimens to Tom Bernatowicz and Kevin Croat of the Washington U. McDonnell
Center for Space Sciences plus Roy Lewis of the U. Chicago Enrico Fermi Institute.
\end{acknowledgments}



\bibliography{ifzx2}

\end{document}